\documentclass[12pt]{article}
\usepackage{color}
\usepackage{subfigure}
\usepackage{epsfig}
\usepackage{amssymb,amsmath}

\begin{document}

\begin{titlepage}
\begin{flushright}
\today
\end{flushright}
\vfill
\begin{centering}

{\bf Constraints for the nuclear parton distributions from $Z$ and $W^\pm$ production at the LHC}

\vspace{0.5cm}
Hannu Paukkunen\footnote{hannu.paukkunen@usc.es} and 
Carlos~A. Salgado\footnote{carlos.salgado@usc.es}

\vspace{1cm}
{Departamento de F\'\i sica de Part\'\i culas and IGFAE, Universidade de Santiago de 
Compostela, Galicia--Spain}

\vspace{1cm} 
{\bf Abstract} \\ 
\end{centering}

The LHC is foreseen to finally bring also the nuclear collisions to the TeV scale thereby
providing new possibilities for physics studies, in particular related to the electro-weak
sector of the Standard Model. We study here the Z and W$^\pm$ production in proton-lead and lead-lead collisions at the LHC,
concentrating on the prospects of testing the factorization and constraining the nuclear modifications of the parton distribution functions (PDFs). Especially, we find that the rapidity asymmetries in proton-nucleus collisions,
arising from the differences in the PDFs between the colliding objects, provide a decisive advantage
in comparison to the rapidity-symmetric nucleus-nucleus case. We comment on 
how such studies will help to improve our knowledge of the nuclear PDFs.

\noindent

\vfill
\end{titlepage}

\setcounter{footnote}{0}

\section{Introduction}
\label{sec:intro}

The production of W$^\pm$ and Z bosons are one of those processes which appear to be well-understood
and under control within the toolkit of perturbative QCD (pQCD). The calculations are already reaching the NNLO
accuracy \cite{Anastasiou:2003ds} with the NNLO correction on top of the NLO one found rather small.
This observation has given confidence to include the recent Tevatron measurements 
\cite{Aaltonen:2009ta,Abazov:2007jy,Aaltonen:2010zza} for  W$^\pm$ and Z production
to the latest global analyses of the free proton parton distribution functions (PDFs)
\cite{Lai:2010vv,Martin:2009iq,Ball:2010de}. These particles will also be copiously produced
at the LHC, and are presumed to provide stringent tests of the Standard Model and eventually
also new constraints to the PDFs.

In contrast to the global analyses of the free proton PDFs \cite{Lai:2010vv,Martin:2009iq,Ball:2010de} the
corresponding nuclear ones \cite{Eskola:2009uj,Schienbein:2009kk,Hirai:2007sx,deFlorian:2003qf} 
suffer from a much more
limited amount and type of experimental data. Consequently, the tests of the factorization --- a working hypothesis
typically assumed in nuclear collisions --- although in general fulfilled, are weaker. The present status
is that one of the sets of nuclear PDFs, EPS09 \cite{Eskola:2009uj}, is able to satisfactorily combine DIS data on nuclear targets, Drell-Yan dilepton production in proton-nucleus collisions and inclusive pion production in
deuteron-gold (dAu) collisions in central rapidities at RHIC with no large tension observed 
among the different data sets. In addition, the same set
provides a good description of neutrino DIS data from nuclear targets \cite{Paukkunen:2010hb}\footnote{See, however, Refs. \cite{Schienbein:2007fs} and \cite{Kovarik:2010uv} for contradicting result.} offering further strength to the hypothesis
of the factorization. All analyses include DIS with nuclei and Drell-Yan production in proton-nucleus collisions, as already done in the pioneering global fits \cite{Eskola:1998iy,Eskola:1998df}, while 
dAu data is only introduced in the EPS-framework, after the EPS08 set in \cite{Eskola:2008ca}.

Along with the unprecedented
center-of-mass energy, the LHC will require a good knowledge of the nPDFs in a kinematical region
never explored before by DIS or other experiments in order to disentangle the physics signals from
the backgrounds. It is therefore of utmost importance that checks of factorization are 
performed also within the LHC-kinematics. One of the goals of the planned proton-nucleus program at
the LHC is precisely to provide the experimental constraints needed for this purpose.
It is worth mentioning that the relevant kinematical region of nPDFs probed by the nuclear program
of the RHIC-BNL was basically already covered by previous experiments, and therefore e.g. the
discovery of the strongly interacting medium in gold-gold collisions there was something that
could not be questioned by the insufficient knowledge of the nPDFs.

In this paper, we study the prospects of observing the nuclear effects --- the difference between
cross-sections in collisions involving heavy ions and those involving only free nucleons --- in
production of heavy bosons at the LHC\footnote{Remarkably, the first Z data in nuclear collisions
have already been published by the ATLAS and CMS collaborations \cite{Collaboration:2010px,Collaboration:2011ua}.}.
We do not separately consider the
different experiments (e.g. cuts imposed on the leptonic transverse
momentum), or centrality dependence (see e.g. \cite{Vogt:2000hp}),
but we will compute the minimum bias rapidity $y_R$ spectra of the decay lepton-pairs in the invariant
mass $M^2$ bin of the heavy boson peak, $M^2 = M_{Z \, {\rm or} \, W^\pm}^2$. In reality, a 
detailed comparison with the experimental results will require an elaborate calculation
with different cuts applied. However, the size of the expected effects
should be well-estimated by the calculations presented here. For example, we have checked that
integrating over the heavy boson resonance profiles or considering charged lepton production instead
of lepton-pairs in the case of W decay would not change the main results.

At the moment of writing this paper, the foreseen nucleon-nucleon center-of-mass energies $\sqrt{s}$ and luminosities
$\mathcal{L}$ in different types of collisions at LHC --- setting the stage for our studies
reported in this paper --- are the following ones:
\begin{equation}
\begin{array}[3]{lcc}
{\rm System} & \sqrt{s} & {\rm Luminosity} \, \mathcal{L}\\
\hline \\
\vspace{0.1cm}
\hspace{-0.5cm} {\rm Proton+Proton} \, {\rm (pp)} & 7 \, {\rm TeV} \quad {\rm and} \quad 14 \, {\rm TeV} & 10^{34} {\rm cm}^{-2}{\rm s}^{-1} \\
\vspace{0.1cm}
\hspace{-0.15cm} {\rm Lead+Lead} \, {\rm (PbPb)} & \hspace{-0.2cm} 2.7 \, {\rm TeV} \quad {\rm and} \quad 5.5 \, {\rm TeV} & 10^{27} {\rm cm}^{-2}{\rm s}^{-1} \\
\hspace{-0.5cm} {\rm Proton+Lead} \, {\rm (pPb)} & 8.8 \, {\rm TeV} & 10^{29} {\rm cm}^{-2}{\rm s}^{-1}
\end{array}
\nonumber
\end{equation}
With clearly smaller nucleon-nucleon luminosities \cite{ref:EPAC2006} and
shorter envisaged running times ($t \sim$ 1 month = $10^6$s), one could worry whether the integrated
yields in pPb- and PbPb-collisions will be large enough to realize the studies presented here.
In the case of pPb collisions the number of expected events would be
$$
\frac{\mathcal{N}^{\rm pPb}}{\Delta y_R \Delta M}
= 208 \times \mathcal{L} \times t \times \left( \frac{\sigma^{\rm pPb}}{\Delta y_R \Delta M} \right) 
\simeq \frac{208}{10} \left( \frac{\sigma^{\rm pPb}}{{\rm pb}} \right).
$$
Where $\sigma^{\rm pPb}$ refers to a cross-section per nucleon --- the ones we will report in the rest of the paper. In the narrow resonance approximation \cite{Anastasiou:2003ds},
$$
\frac{\mathcal{N}^{\rm pPb}}{\Delta y_R} = \frac{\pi \Gamma_V}{2} \times \frac{\mathcal{N}^{\rm pPb}}{\Delta y_R \Delta M} \Big|_{M=M_V},
$$
where $\Gamma_V$ is the width of the heavy boson. As an example, for the Z-production this translates
(see Fig.~\ref{Fig:Z_pPb_Spectrum} ahead) to around 4000 events in the mid-rapidity $y_R=0$ unit bin and still
around 1000 events at $|y_R| = 3$ unit bin. Moreover, knowing that larger luminosities could be possible
\cite{d'Enterria:2009er}, the heavy boson study in proton-nucleus collisions at the LHC looks feasible.
For PbPb-collisions the corresponding numbers are similar as the smaller luminosity
is compensated by a additional factor of 208.

In what follows, we first describe the ingredients of our calculations, and then
present the results for pPb and PbPb collisions for Z and W$^\pm$ bosons respectively.
The last section summarizes the main results.

\section{Framework of the Calculations}
\label{sec:framework}

\subsection{Z-Production}
\label{sec:Z-Production}
The reaction  we consider is the Drell-Yan production of lepton pairs with invariant mass $M^2$ and rapidity $y_R$ (corresponding to the invariant mass and rapidity of the force-carrying boson in the absence of electroweak final state radiation), in a collision of two nucleons ${\rm H}_1$ and ${\rm H}_2$ at high center-of-mass energy $\sqrt{s}$:
\begin{equation}
 \frac{d^2\sigma \left( \sqrt{s}, {\rm H}_1 + {\rm H}_2 \rightarrow \ell^+ + \ell^- + {\rm X} \right)}{dM^2dy_R} \label{eq:Z_LO}.
\end{equation}
The framework of the calculations is the NLO-level pQCD improved parton model.
The relevant expressions can be found for example from \cite{Anastasiou:2003ds} or obtained
from \cite{Paukkunen:2009ks,Sutton:1991ay} with obvious change of couplings indicated below. In order to illustrate
the ingredients and to exactly specify the computed quantity, we record here the leading order part of
the cross-sections. For this particular process it reads
\begin{equation}
 \frac{d^2\sigma}{dM^2dy_R} = \frac{4\pi\alpha_{\rm em}^2}{9sM^2} \sum_q c_q^2
 \left[ q^{(1)}(x_1,Q_f^2)\overline{q}^{(2)}(x_2,Q_f^2) + \overline{q}^{(1)}(x_1,Q_f^2) {q}^{(2)}(x_2,Q_f^2) \right],
\end{equation}
where $q^{(1)}(x_1,Q_f^2)$ and $q^{(2)}(x_2,Q_f^2)$ are the PDFs of the colliding nucleons read at momentum fractions $x_{1,2} \equiv (M / \sqrt{s}) e^{y_R,\, -y_R}$, and factorization scale $Q_f^2$. In this work we always choose $Q_f^2 = M^2$. The electroweak couplings $c_q^2$ can be expressed as
\begin{eqnarray}
c_q^2 \equiv e_q^2 & - & 2 \chi e_q \frac{M^2\left( M^2-M_Z^2\right)}{(M^2-M_Z^2)^2 + M_Z^2 \Gamma_Z^2} V_\ell V_q \label{eq:coup} \\
 & + & \chi^2 \frac{M^4}{(M^2-M_Z^2)^2 + M_Z^2 \Gamma_Z^2} \left( A_\ell^2 + V_\ell^2 \right)\left( A_q^2 + V_q^2 \right), \nonumber  \\
 \\ \nonumber
& & \hspace{-2.15cm} \chi \equiv (\sqrt{2}G_F M_Z^2)/(16\pi \alpha_{\rm em}),
\end{eqnarray}
where $\Gamma_Z = 2.4952\, {\rm GeV}$ is the total width of the Z-boson, $\alpha_{\rm em} = 1/137.03604$ the QED coupling constant, $G_F = 1.16637 \times 10^{-5} \, {\rm GeV}^{-2}$ the Fermi constant, and $M_Z = 91.1876\, {\rm GeV}$ the mass of the Z-boson. The first term in Eq.~(\ref{eq:coup}) involving only the quark charge squared $e_q^2$, is the pure QED part corresponding to the one virtual photon exchange. The third term origins correspondingly from the pure Z-exchange, and the second one is the Z$\gamma$-interference term. The vector and axial couplings are
\begin{eqnarray}
V_\ell & = & -1 + 4 \sin^2 \theta_{\rm W}, \hspace{1.25cm} A_\ell  =  -1 \nonumber \\
V_{q_u} & = & 1 - \frac{8}{3} \sin^2 \theta_{\rm W}, \hspace{1.3cm} A_{q_u}  =  1 \\
V_{q_d} & = & -1 + \frac{4}{3} \sin^2 \theta_{\rm W}, \hspace{1cm} A_{q_d}  =  -1 \nonumber,
\end{eqnarray}
for lepton $\ell$, up-type quarks $q_u = u, c$, and down-type quarks $q_u = d, s, b$.
The $\sin^2 \theta_{\rm W} = 0.23143$ is the weak-mixing angle.

Let us note, that since we are mostly interested in certain ratios of the cross-sections, the sensitivity to the specific choice of the numerical values of the couplings indicated above, is not that critical. Also, the effect of fixing the factorization scale $Q_f^2$ to the invariant mass of the lepton pair, does not bear large consequences to the cross-section ratios --- especially not to the nuclear effects --- as the $Q^2$-dependence of the nuclear effects is rather mild at such large values of $Q^2$.

\subsection{W$^\pm$-Production}
\label{sec:W-Production}
In the case of W$^\pm$-production, we consider the leptonic channel
\begin{equation}
 \frac{
 d^2\sigma \left( \sqrt{s}, 
 {\rm H}_1 + {\rm H}_2 \rightarrow \left\{
 \begin{array}{cc}
  \ell^+ + \nu \\ \ell^- + \overline \nu
 \end{array} \right.
  + {\rm X} \right)}{dM^2dy_R}, \label{eq:W_process}
\end{equation} 
assuming that the missing neutrino momentum could be reconstructed by the experiment. In reality, extracting the rapidity spectrum for W$^\pm$ is not possible without some theory input  \cite{Aaltonen:2009ta,Lohwasser:2010sp} as the neutrino momentum cannot be fully reconstructed. In this respect the single lepton cross-section $d^2\sigma / (dy_R^{\ell^\pm}dp_T^{\ell^\pm})$ for ${\rm H_1 + H_2 \rightarrow \ell^\pm + X}$ should be more useful as a better defined quantity from the experimental point of view. 

However, at the lepton transverse momentum $p_T^{\ell^\pm} \approx M_W/2$ --- where the contamination of the
charged lepton spectra from heavy quark decays should be negligible --- it can be argued (see the Appendix) that
the ${\rm H_1 + H_2 \rightarrow \ell^\pm + X}$ cross-section is sensitive to the same partonic combination
as the ${\rm H_1 + H_2 \rightarrow \ell^\pm + \nu + X}$ process. Indeed, we have checked that the main results
for the $\ell^\pm$ production at $p_T^{\ell^\pm} \approx M_W/2$ are essentially unchanged with respect to the $W^\pm$ production at $M=M_W$ considered here.  As demonstrated in the Appendix, even if a wider $p_T^{\ell^\pm}$-window
is utilized, the main results are still well-estimated by the calculations presented here.

\begin{figure}[!htb]
\center
\includegraphics[scale=0.9]{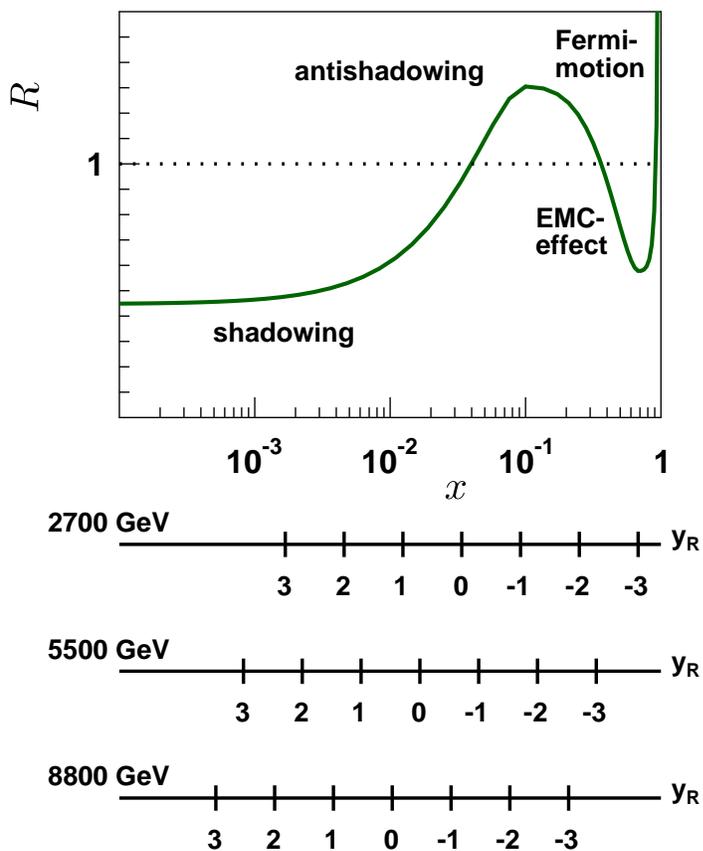}
\caption[]{\small The typical shape of the nuclear modifications as a function of $x$, and the kinematical reach for Z-production at three different center-of-mass energies and $M^2=M_Z^2$ for some values of rapidity.}
\label{Fig:kin}
\end{figure}

In leading order, the cross-section for the process of Eq.~(\ref{eq:W_process}) reads
\begin{eqnarray}
 \frac{d^2\sigma^{W^\pm}}{dM^2dy_R} & = & \frac{\pi \alpha_{\rm em}^2}{36 s M^2 \sin^4 \theta_{\rm W}}
 \frac{M^4}{\left( M^2-M_W^2 \right)^2 + M_W^2 \Gamma_W^2}  \\
 & \times & \sum \left|V_{ij} \right|^2 \left[ q_i^{(1)}(x_1,Q_f^2)\overline{q}_j^{(2)}(x_2,Q_f^2) + \overline{q}_j^{(1)}(x_1,Q_f^2) {q_i}^{(2)}(x_2,Q_f^2) \right], \nonumber
\end{eqnarray}
where $V_{ij}$s denote the elements of the CKM-matrix, and 
$$
\begin{array}{llc}
i = u, c     & j = d, s, b & {\rm for \,\, W^+} \\ 
i = d, s, b  & j = u, c    & {\rm for \,\, W^-}. 
\end{array}
$$
We take the W mass to be $M_W = 80.398 \, {\rm GeV}$, the total width $\Gamma_W = 2.141 \, {\rm GeV}$, and set \cite{Nakamura:2010zzi}
\begin{eqnarray}
 |V_{\rm ud}| & = & 0.974 \hspace{2cm} |V_{\rm us}| = 0.225  \nonumber \\
 |V_{\rm cd}| & = & 0.225  \hspace{2cm}  |V_{\rm cs}| = 0.973 \\  
 |V_{\rm cb}| & = & 0.041  \hspace{2cm}  |V_{\rm ub}| = 0.0035. \nonumber
\end{eqnarray}

\subsection{Nuclear PDFs}
\label{sec:Nuclear_PDFs}

The usual way of constructing a set of nuclear PDFs is to take a set of free proton PDFs on top of which nuclear modifications are implemented. 
The EPS09 set  \cite{Eskola:2009uj}, employed for the calculations in this paper and which allows for a proper error analysis, has been obtained using CTEQ6.1M \cite{Stump:2003yu} as the reference proton distributions. Here we will use instead CTEQ6.6 \cite{Nadolsky:2008zw} which includes a more elaborated prescription for the treatment of the heavy quarks\footnote{We have checked 
that the change in the nuclear ratios EPS09 or the error analysis are not very much affected if the global fit is performed by taking 
CTEQ6.6 as the reference proton set.}. To be precise, the PDFs used to compute cross-sections involving heavy ions are
\begin{equation}
 f^A = \frac{Z}{A}f^{{\rm proton}, A} + \frac{N}{A}f^{{\rm neutron}, A},
\end{equation}
where $Z$ and $N$ are the number of protons and neutrons in a nucleus with mass number $A$. The bound proton and neutron PDFs are denoted by $f^{{\rm proton}, A}$ and $f^{{\rm neutron}, A}$,  the latter ones obtained from those of a bound proton by a basis of the isospin symmetry,
\begin{equation}
 d^{{\rm neutron}, A} = u^{{\rm proton}, A}, \qquad
 u^{{\rm neutron}, A} = d^{{\rm proton}, A}.
\end{equation}
For other flavors the bound proton and bound neutron PDFs are equal. In EPS09, the nuclear modifications to the PDFs are defined through multiplicative, scale and flavor dependent factors $R_f(x,Q^2)$:
\begin{equation}
 f^{{\rm proton}, A}(x,Q^2) \equiv R_f(x,Q^2) f^{{\rm free \, proton}}(x,Q^2).
\end{equation}
As a concrete example, the down valence quark distribution $d_{\rm V} \equiv d - \overline d$ would be
\begin{equation}
 d^A_{\rm V} = \frac{Z}{A} R_{d_V} d_{\rm V}^{{\rm free \, proton}} + \frac{N}{A} R_{u_V} u_{\rm V}^{{\rm free \, proton}}.
\end{equation}

To understand the shapes of the calculated nuclear modifications presented in the following sections, we plot
the typical modifications of the corresponding PDFs $R_f(x,Q^2)$ in Figure~\ref{Fig:kin}. This figure also indicates the
values of momentum fractions $x_2$ that enter to the leading order term of Z-production Eq.~(\ref{eq:Z_LO}) (they
are also the smallest $x_2$-values involved in the NLO-level calculations involving integrations) at three different
center-of-mass energies foreseen to be realized at the LHC nuclear program.

In what follows, we will especially consider the uncertainties related to the PDFs and to their nuclear modifications. To this end, EPS09 and CTEQ6.6 contain error sets $S_k^\pm$ of PDFs forming an inseparable part of the published best-fits $S_0$. Here, we use the symmetric prescription
\begin{equation}
 \Delta X = \frac{1}{2} \sqrt{ \sum_k \left[ X(S_k^+)-X(S_k^-) \right]^2 },
\end{equation}
to calculate the uncertainty related to the quantity $X$, which could be either a cross-section or a combination of them.
If the experimental results end up being more accurate than the predicted uncertainty bands calculated in this way, the data will be evidently capable to set useful constraints to the PDFs.

The error analysis carried out in this manner provides a propagation of the experimental uncertainties 
included in the corresponding global fit of 
PDFs. It is, however, known that these global fits can be biased by the assumptions in the initial conditions to be used in the
DGLAP evolution equations. These biases can be made smaller when the amount of data is large.  In the nuclear
case the two main biasing assumptions at the parametrization scale $Q^2=Q_0^2$ are:

\begin{enumerate}

\item
The flavor decomposition is, at the moment, very badly known: EPS09 assumed that
$R_{u_V}(x,Q_0^2) = R_{d_V}(x,Q_0^2)$ and $R_{\overline u}(x,Q_0^2) = R_{\overline d}(x,Q_0^2) = R_{\overline s}(x,Q_0^2)$, at the parametrization scale. Although this feature to some extent disappears in $Q^2$-evolution to the scales of heavy boson mases, we anticipate that relaxing this assumption would especially increase the uncertainties of W$^\pm$ cross-sections, being more sensitive to the flavor decomposition.

\item
 At the moment, the experimentally constrained $x$-region is limited to $x \gtrsim 10^{-2}$. Below,
the shape of the nuclear modifications and the size of their uncertainties are  restricted by the assumed form of the fit function. Such a
perversion makes the uncertainties to appear smaller than they actually are. For pPb collisions at
$\sqrt{s} = 8.8 \, {\rm TeV}$ this would happen right outside the midrapidity $y_R > 0$, as can be
read off from the Figure~\ref{Fig:kin}. That is, data precise enough would also facilitate having more
small-$x$ freedom in the fit functions thereby obtaining more realistic estimations for the uncertainties in nPDFs.

\end{enumerate}

In summary, although the uncertainties presented in the following are the best estimates available for the propagation of present experimental information to the unknown LHC regime, they are to be taken as lower limits of the real uncertainties.

\section{Results}
\label{sec:Results}

\subsection{Z Production in p+Pb collisions}
\label{sec:Z_pPb}

\begin{figure}[!htb]
\center
\includegraphics[scale=0.5]{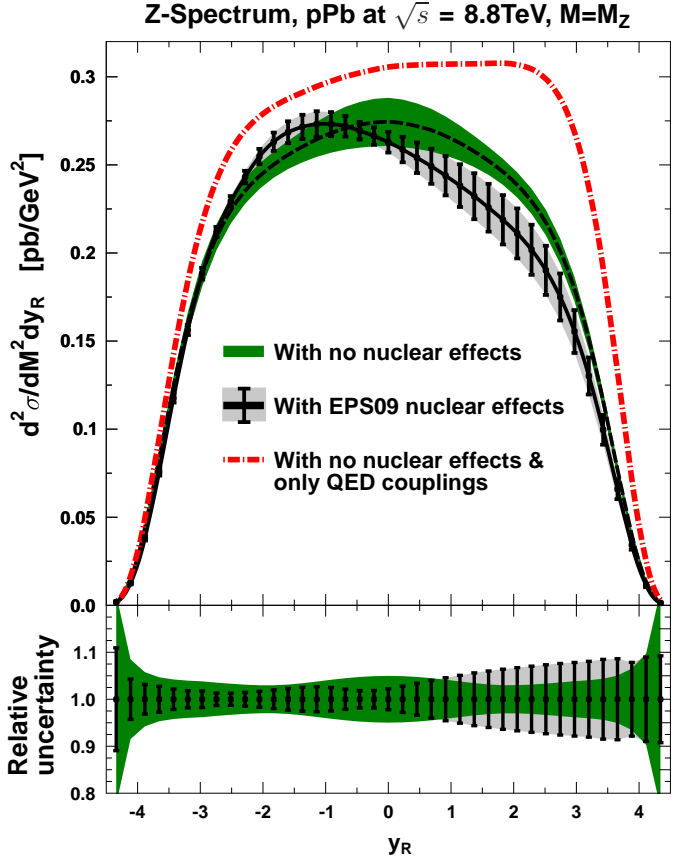}
\caption[]{\small The calculated cross-sections for Z-boson production in pPb collisions at $\sqrt{s} = 8.8 \, {\rm TeV}$ at the Z-pole $M^2=M_Z^2$. The dashed line represents the central prediction calculated without applying the nuclear corrections to PDFs, and the green band is the uncertainty range derived from CTEQ6.6. The solid line is the prediction computed by CTEQ6.6 applying the nuclear effects from EPS09. The error bars quantify the uncertainty derived from EPS09 uncertainty sets. The red dashed-dotted curve is the prediction with only QED couplings (first term in Eq.~\ref{eq:coup}) multiplied by 1100 with no nuclear corrections to PDFs.  The lower panel shows the relative uncertainties with the same color codes.}
\label{Fig:Z_pPb_Spectrum}
\end{figure}
\begin{figure}[!htb]
\center
\includegraphics[scale=0.5]{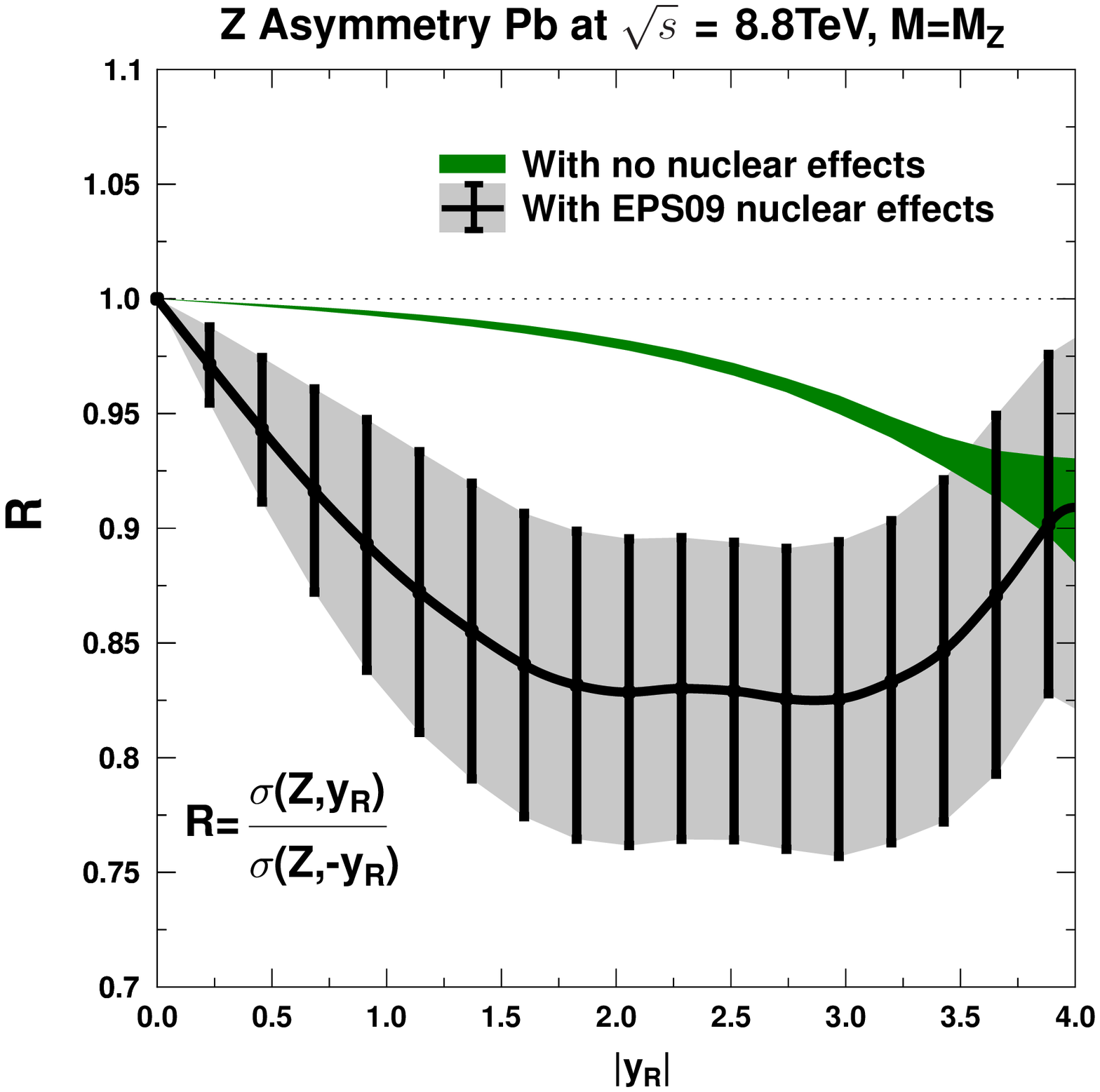}
\caption[]{\small 
The predicted ratio between forward and backward yields of Z bosons in pPb collisions at $\sqrt{s} = 8.8 \, {\rm TeV}$ at the Z-pole, $M^2=M_Z^2$. The green solid band represents the prediction calculated with CTEQ6.6 without applying the nuclear effects, and the solid black barred line with gray shade is the prediction computed by CTEQ6.6 applying the nuclear effects from EPS09. The error bars quantify the uncertainties resulting from the EPS09 uncertainty sets.}
\label{Fig:Z_pPb_Asym}
\end{figure}

The usefulness of Z-production in pPb collisions lies in the almost symmetric rapidity spectrum predicted in the absence of nuclear effects in PDFs. This is demonstrated in Figure~\ref{Fig:Z_pPb_Spectrum}, where the dashed line and the green band represent the prediction and its uncertainty without nuclear effects in PDFs. The small departure from a totally symmetric spectrum is only due to the different relative content of $u$ and $d$ quarks in the proton and the lead nuclei. The
evident smallness of such \emph{isospin} effect can be traced back to the electroweak couplings. Especially since
\begin{equation}
 \frac{\sigma_{y_R}}{\sigma_{-y_R}} - 1 \propto \left( \frac{N}{A} \right) \left( \frac{c_u^2 - c_d^2}{c_u^2 + c_d^2} \right),
\end{equation}
and
\begin{eqnarray}
 \frac{c_u^2 - c_d^2}{c_u^2 + c_d^2} & \approx & 0.13, \hspace{0.5cm} {\rm when} \hspace{0.5cm} M^2 = M_Z^2 \\
 \frac{c_u^2 - c_d^2}{c_u^2 + c_d^2} & \approx & 0.60, \hspace{0.5cm} {\rm when} \hspace{0.5cm} M^2 \ll M_Z^2,
\end{eqnarray}
the asymmetry between forward and backward rapidities is suppressed close to the Z-peak with respect to photon exchange.
This is demonstrated by the red dashed-dotted line in Figure~\ref{Fig:Z_pPb_Spectrum}, which represents the calculation with only QED-couplings (normalized by a factor of 1100 for clarity)  bringing clearly out the effects associated to the different isospin content of the proton and the lead nucleus.

Applying the nuclear effects to the PDFs, however, significantly modifies the almost symmetric distribution and the induced difference is clearly visible by eye already in the level of absolute spectrum. This is evident from Figure~\ref{Fig:Z_pPb_Spectrum}, where the solid barred line represents the calculation with the EPS09 nuclear 
effects in PDFs, and the error bars around this line standing for its errors. 

Although the uncertainty due to free proton PDFs in the absolute spectrum in Figure~\ref{Fig:Z_pPb_Spectrum} is clearly non-negligible, the absence of strong isospin effects in the Z-spectrum in pPb collisions can be exploited to cleanly extract the nuclear corrections to PDFs. This can be done by simply considering the ratio between Z-production in forward and backward directions
\begin{equation}
 \frac{d^2\sigma^{Z, y_R}}{dM^2dy_R} / \frac{d^2\sigma^{Z, -y_R}}{dM^2dy_R}, \label{eq::Z_Ratio}
\end{equation}
in which the uncertainties due to free proton PDFs tend to cancel to large extent, thanks to the symmetry property mentioned earlier. This is demonstrated by the very thin green band in Figure.~\ref{Fig:Z_pPb_Asym}. On the other hand, the asymmetry and its uncertainty originating from the nuclear effects is reflected as a wide barred gray band in Figure~\ref{Fig:Z_pPb_Asym}. The reason for a rather large nuclear effect is clear: at forward rapidity ($y_R > 0$) one is more sensitive to the small-$x$ shadowing ($R_i < 1$), whereas at backward direction ($y_R < 0$) the antishadowing ($R_i > 1$) sets in, see Figure~\ref{Fig:kin}. In the ratio of Eq.~(\ref{eq::Z_Ratio}) these two effects nicely add pronouncing the predicted nuclear effect.

\begin{figure}[!htb]
\center
\includegraphics[scale=0.4]{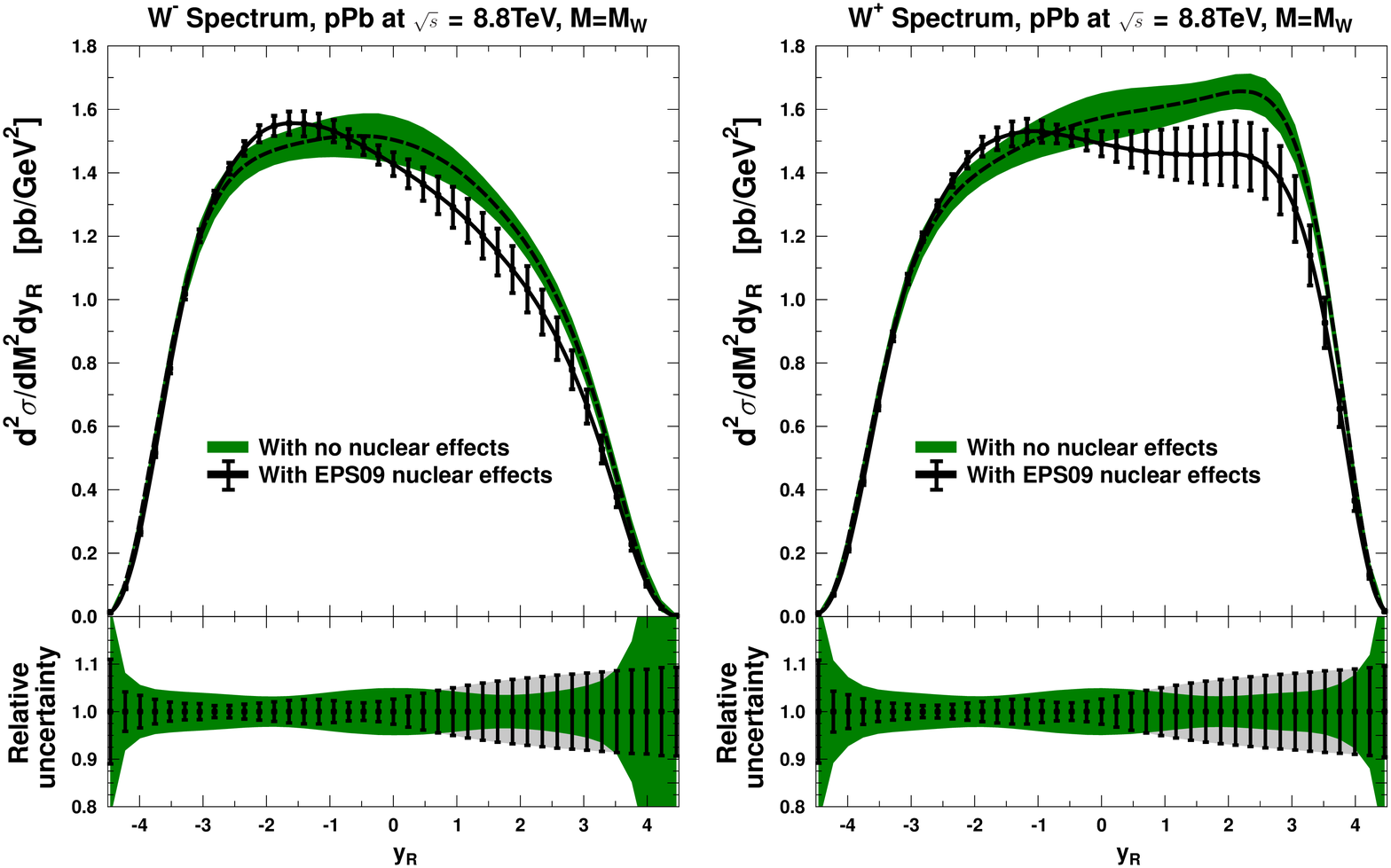}
\caption[]{\small  The calculated cross-sections for W$^\pm$-production in pPb collisions at $\sqrt{s} = 8.8 \, {\rm TeV}$ at the W-pole, $M^2=M_W^2$. The dashed line represents the central prediction calculated with CTEQ6.6 without applying the nuclear effects, and the green band is the uncertainty range derived from CTE6.6 PDFs. The solid line is the prediction computed by CTEQ6.6 applying the nuclear effects from EPS09. The error bars quantify the uncertainties resulting from the EPS09 uncertainty sets.  The lower panels show the relative uncertainties with the same color codes.}
\label{Fig:W_pPb_Spectrum}
\end{figure}

\subsection{W$^\pm$ Production in p+Pb collisions}
\label{sec:W_pPb}

The extraction of the pure nuclear effects from ${\rm W}^\pm$-production is not that straightforward than isolating them from the Z-spectra. This is because the symmetry present in Z-spectra in the absence of nuclear effects in PDFs does not carry over to W$^\pm$-production as can clearly be seen from the asymmetric green bands in Figure~\ref{Fig:W_pPb_Spectrum} where the W$^\pm$-spectra in pPb-collisions are plotted. As earlier, the predictions with and without the nuclear effects in PDFs are plotted separately. 

\begin{figure}[!htb]
\center
\includegraphics[scale=0.37]{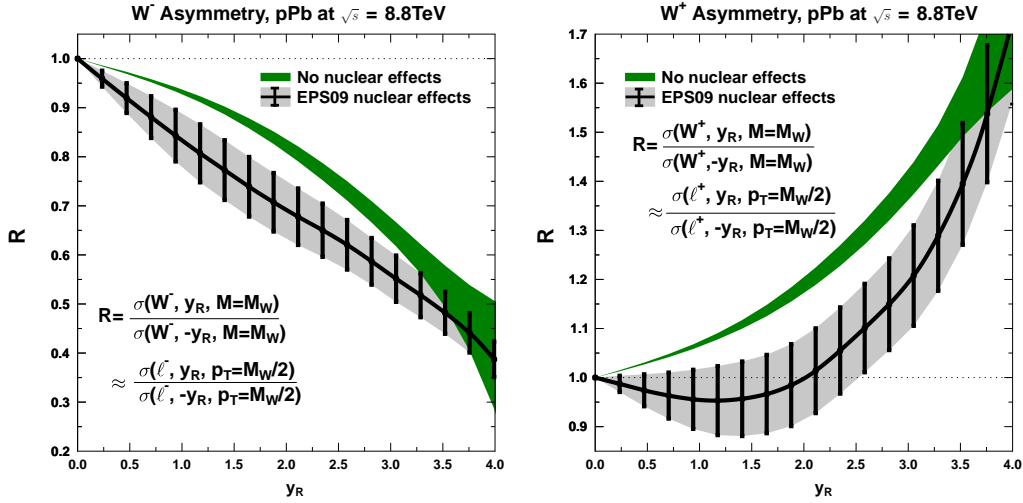}
\caption[]{\small 
The predicted ratio between forward and backward yields of and W$^+$ (right panel) and and W$^-$ (left panel) bosons in pPb
collisions at $\sqrt{s} = 8.8 \, {\rm TeV}$. The green solid band represents the prediction calculated with CTEQ6.6 without applying the nuclear effects, and the solid black barred line with gray shade is the prediction computed by CTEQ6.6 applying the nuclear effects from EPS09. The error bars quantify the uncertainties resulting from the EPS09 uncertainty sets.}
\label{Fig:W_pPb_Asym1}
\end{figure}
\begin{figure}[!htb]
\center
\includegraphics[scale=0.37]{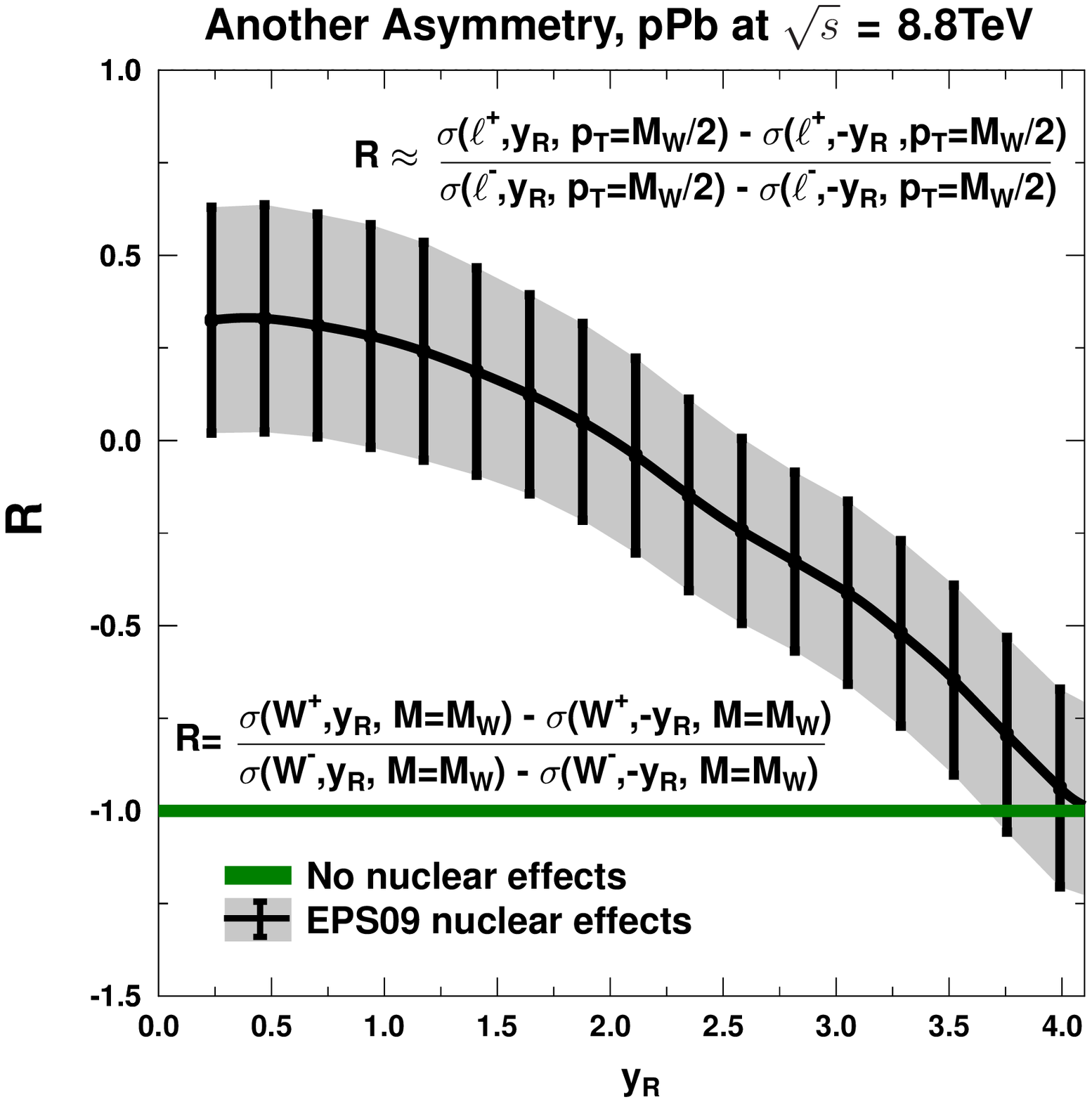} 
\includegraphics[scale=0.37]{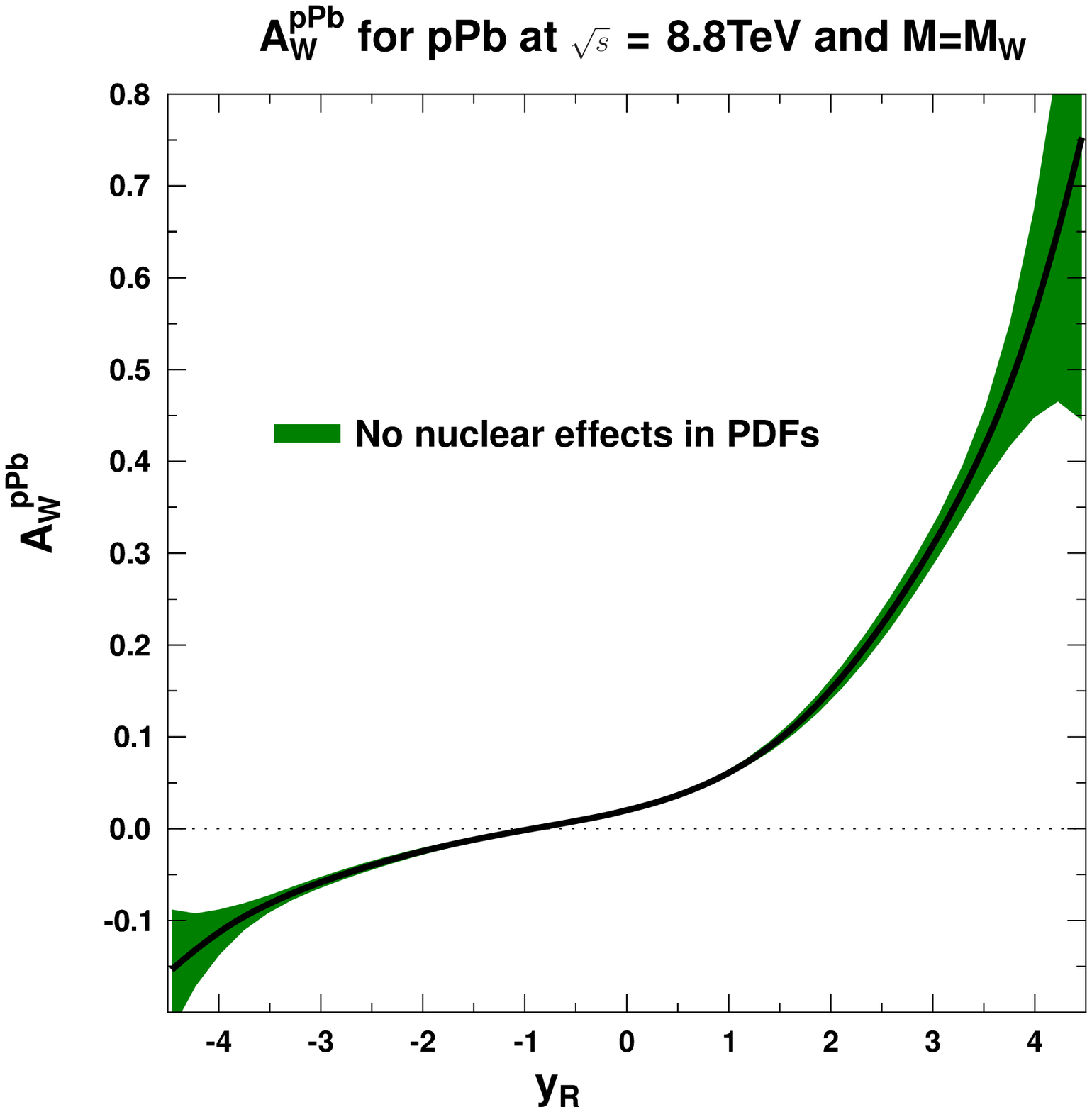}
\caption[]{\small The predicted ratio between forward-backward difference of W$^+$ and and W$^-$ bosons (left) and
the W charge asymmetry (right) in pPb collisions at $\sqrt{s} = 8.8 \, {\rm TeV}$. The green solid line represents the prediction calculated with CTEQ6.6 without applying the nuclear effects. The solid black barred line is the prediction computed by CTEQ6.6 applying the nuclear effects from EPS09. The error bars quantify the uncertainties resulting from the EPS09 uncertainty sets (not visible for the W charge asymmetry).}
\label{Fig:W_pPb_Asym2}
\end{figure}

In analogy to the Z-production we first consider, in Figure~\ref{Fig:W_pPb_Asym1}, the ratios
\begin{equation}
 \frac{d^2\sigma^{W^\pm, y_R}}{dM^2dy_R} / \frac{d^2\sigma^{W^\pm, -y_R}}{dM^2dy_R}.
\end{equation}
Due to the asymmetry present already without nuclear effects in PDFs, the green bands denoting the CTEQ6.6 range deviate significantly from unity and tend to be wider than in the case of Z-production. As the uncertainty range from EPS09 is generally larger, one could use these ratios --- if they are measured at LHC --- as an input to a nPDF fit.
There is, however, another combination of W$^\pm$ cross-sections that is much more inert to the underlying set of free proton PDFs and its uncertainties, namely
\begin{equation}
 \left[ \frac{d^2\sigma^{W^+,y_R}}{dM^2dy_R} - \frac{d^2\sigma^{W^+, -y_R}}{dM^2dy_R} \right] / \left[ \frac{d^2\sigma^{W^-, y_R}}{dM^2dy_R} - \frac{d^2\sigma^{W^-, -y_R}}{dM^2dy_R} \right]. \label{eq:Wasym1}
\end{equation}
This quantity is plotted in Figure~\ref{Fig:W_pPb_Asym2}, revealing how already at midrapidity, where the cross-sections should be well measurable, there should be a large effect emerging from nuclear modifications to the PDFs. In fact,
utilizing the leading order formulas, one may easily show that in the absence of nuclear modifications in PDFs this
quantity is exactly unity and the deviations from unity are solely due to the nuclear effects in the PDFs.

We note here that the usual W charge asymmetry defined as
\begin{equation}
 A_W(y_R) \equiv \left[ \frac{d^2\sigma^{W^+,y_R}}{dM^2dy_R} - \frac{d^2\sigma^{W^-, y_R}}{dM^2dy_R} \right] / \left[ \frac{d^2\sigma^{W^+, y_R}}{dM^2dy_R} + \frac{d^2\sigma^{W^-, y_R}}{dM^2dy_R} \right],
\end{equation}
is much more sensitive to the free proton PDFs than to their nuclear modifications.
This insensitivity is illustrated in Figure~\ref{Fig:W_pPb_Asym2} for pPb collisions where
the nuclear modification to the curve is not visible. Such insensitivity to the nuclear 
modifications can be partly related to the assumptions made in EPS09 (see Sec.~\ref{sec:Nuclear_PDFs})
about the flavor-blindess of the nuclear effects at the parametrization scale. At the
moment it is unfortunately difficult to estimate how much larger the nuclear uncertainties would
grow if this assumption would have been released. However, in the case the nuclear uncertainty proves
truly negligible, the quantity $A_W$ in pPb collisions could turn out surprisingly useful for
the studies of the free proton PDFs.

\subsection{Z and W$^\pm$ production in Pb+Pb collisions}
\label{sec:Z_PbPb}

\begin{figure}[!htb]
\center
\includegraphics[scale=0.35]{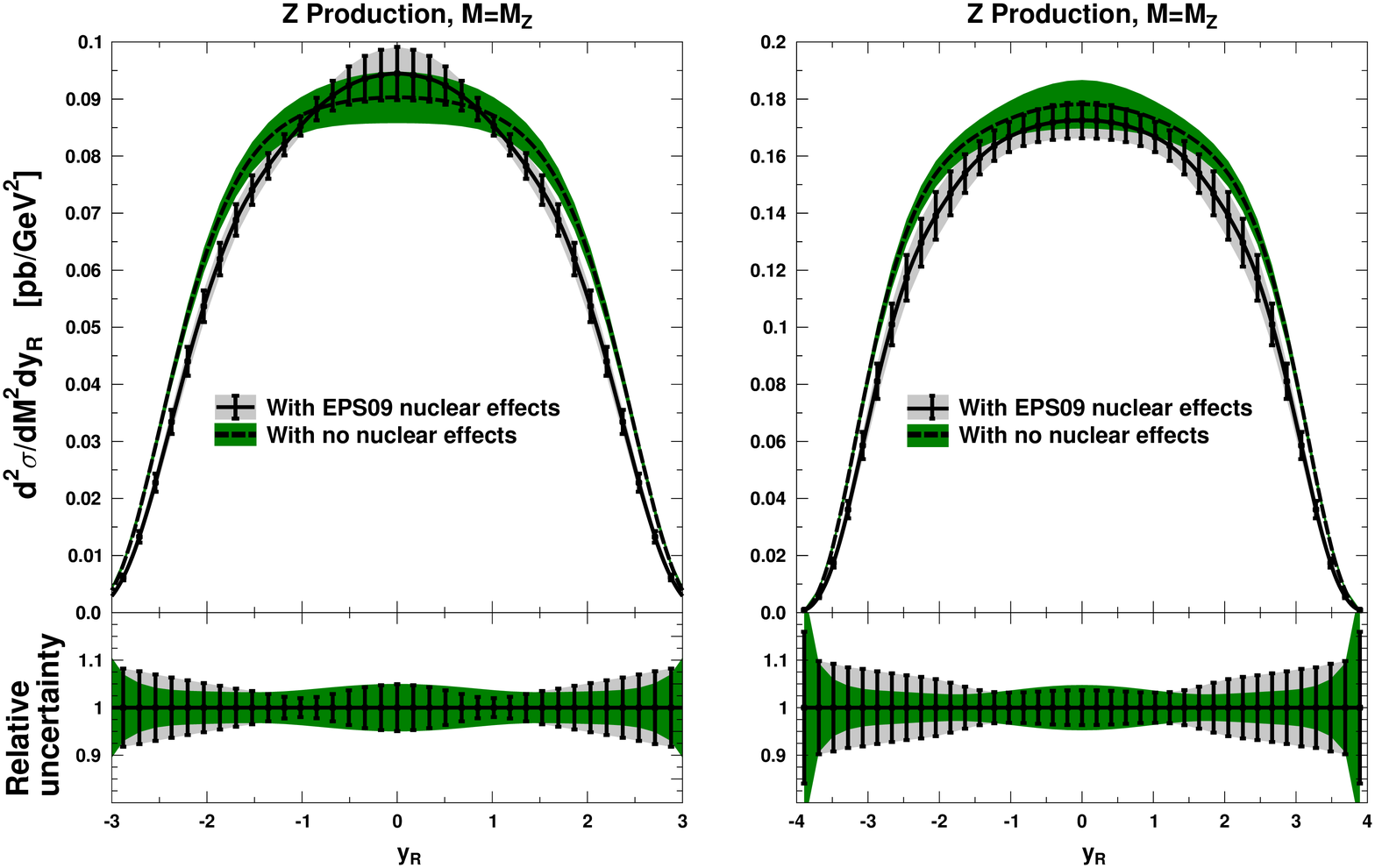}
\includegraphics[scale=0.36]{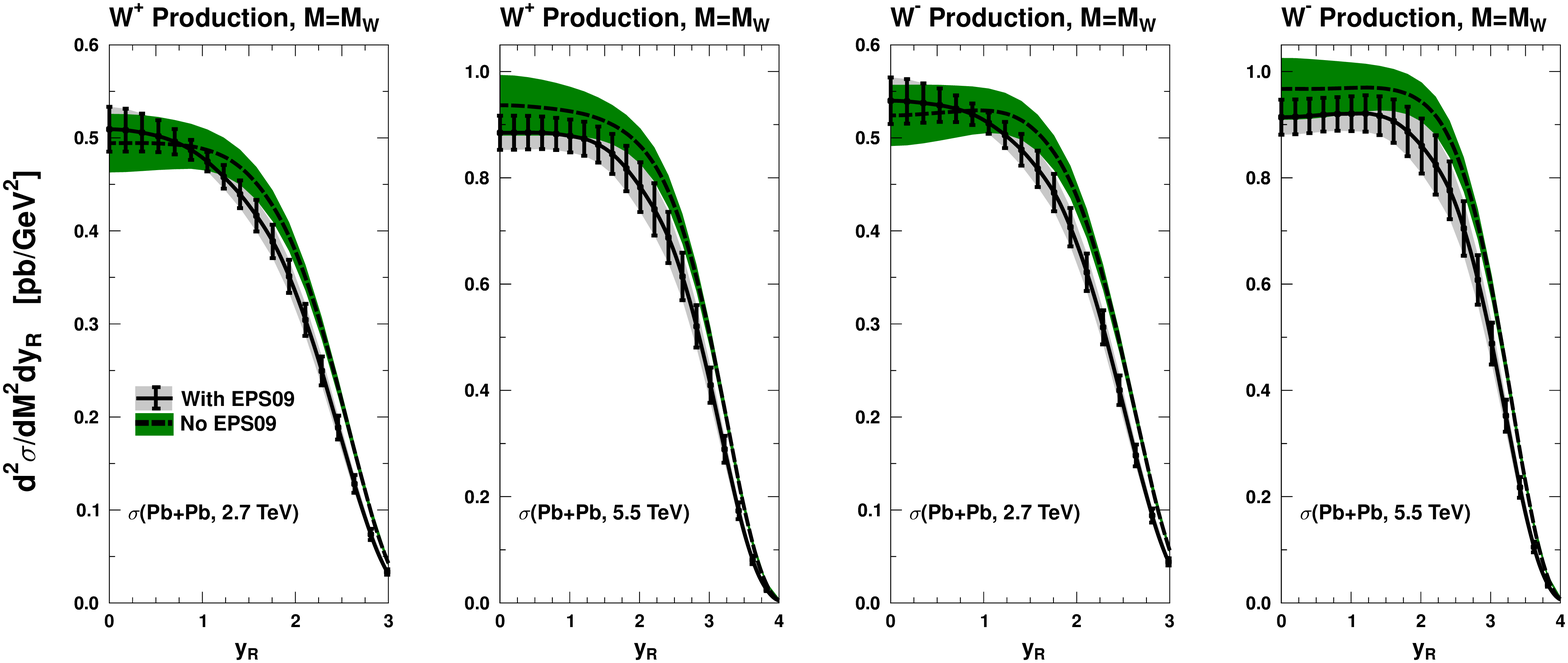}
\caption[]{\small
The predicted Z and $W^\pm$ rapidity spectra in PbPb collisions 
at the heavy boson pole, $M^2=M_{\rm Z,W^\pm}^2$ with $\sqrt{s}=2.7\,{\rm TeV}$ and $\sqrt{s}=5.5\,{\rm TeV}$. The green solid band represents the prediction calculated with CTEQ6.6 without applying the nuclear effects, and the solid black barred line with gray shade is the prediction computed by CTEQ6.6 applying the nuclear effects from EPS09. The error bars quantify the uncertainties resulting from the EPS09 uncertainty sets.}
\label{Fig:Z_PbPb}
\end{figure}

Before the realization of any pPb-runs, the LHC will be operated in pp- and PbPb-modes. Therefore, it is relevant to check if already heavy boson production in PbPb-collisions would provide useful information about the nuclear effects in PDFs. In spite of the dense QCD-matter being created in PbPb-collisions, the leptons from the decays of heavy bosons should penetrate practically unaffected through this medium (for a short summary, see \cite{ConesadelValle:2009vp}) justifying the interpretation based on the pQCD parton model.

First, in Figure~\ref{Fig:Z_PbPb}, we plot the predicted absolute rapidity spectrum for Z and $W^\pm$ bosons
in Pb+Pb collisions. The total uncertainty (EPS09 and CTEQ6.6 errors added in quadrature) there is
typically less than 10\%. This uncertainty is of the same order as the quoted uncertainty when 
the cross-section is extracted from the absolute yields utilizing a relation \cite{d'Enterria:2003qs}
\begin{equation}
 \frac{d^2N^{\rm PbPb}}{dM^2dy_R} = \frac{\langle N_{\rm coll}\rangle}{\sigma_{\rm NN}^{\rm inelastic}} \frac{d^2\sigma^{\rm PbPb}}{dM^2dy_R},
\end{equation}
where $\sigma_{\rm NN}^{\rm inelastic}$ is the total inelastic nucleon-nucleon cross-section and
$\langle N_{\rm coll}\rangle$ is the average number of binary nucleon-nucleon collisions extracted
from a Glauber model Monte Carlo simulation. This indicates that the theory calculations appear already
accurate enough to serve as an independent check of the Glauber model.

\begin{figure}[!htb]
\center
\includegraphics[scale=0.35]{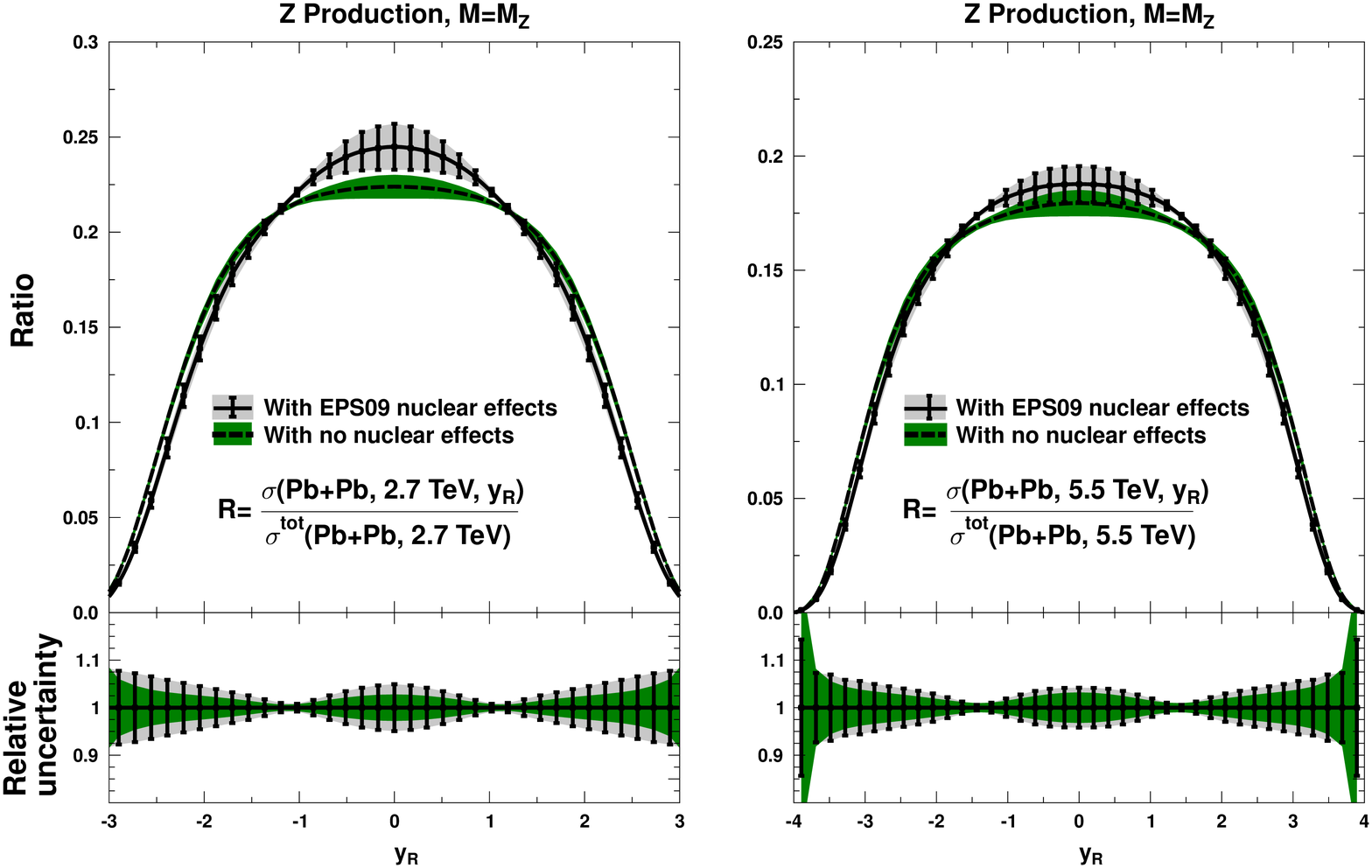}
\includegraphics[scale=0.36]{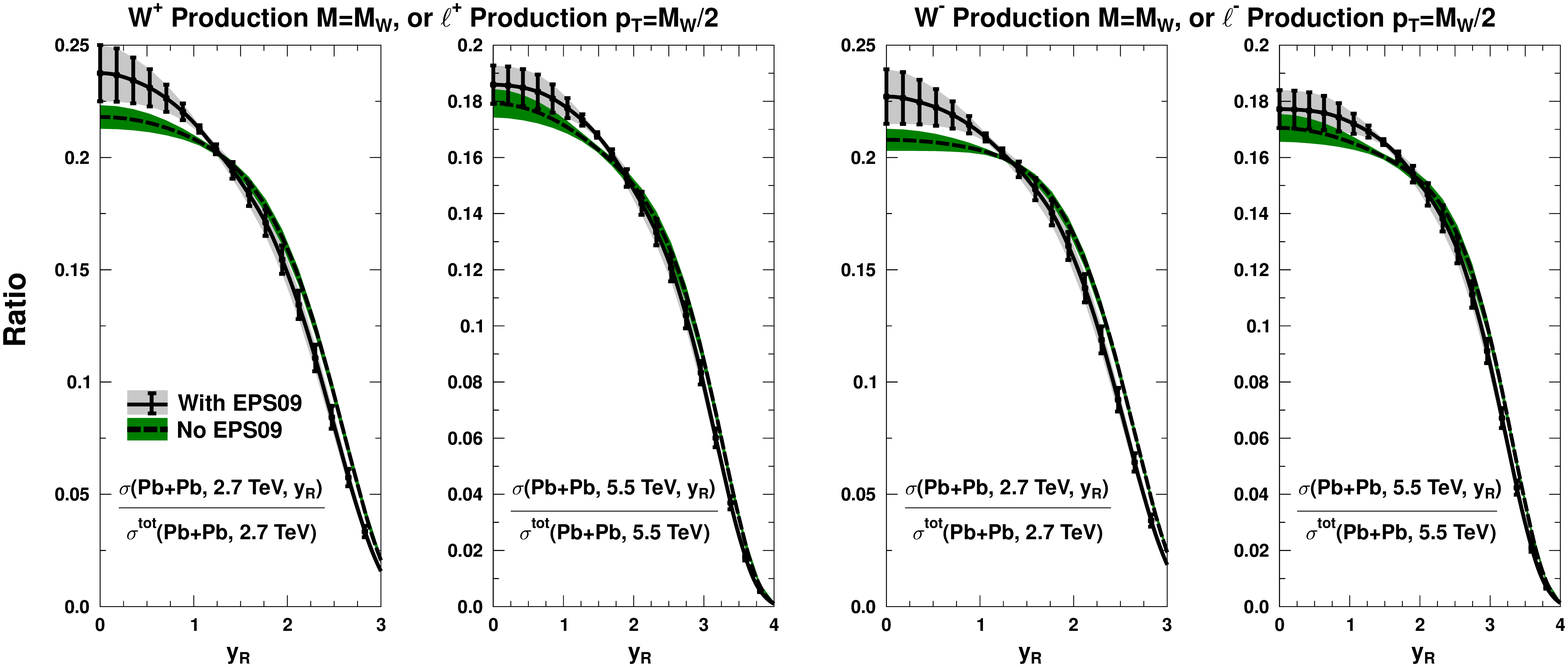}
\caption[]{\small
The predicted Z and $W^\pm$ rapidity spectra in PbPb collisions normalized to the integrated
cross-section at the heavy boson pole, $M^2=M_{\rm Z,W^\pm}^2$ ($p_T=M_W/2$ for charged lepton production) with $\sqrt{s}=2.7\,{\rm TeV}$ and $\sqrt{s}=5.5\,{\rm TeV}$. The green solid band represents the prediction calculated with CTEQ6.6 without applying the nuclear effects, and the solid black barred line with gray shade is the prediction computed by CTEQ6.6 applying the nuclear effects from EPS09. The error bars quantify the uncertainties resulting from the EPS09 uncertainty sets. The symbol $\sigma^{\rm tot}$ refers to the cross-section integrated over the rapidity.}
\label{Fig:Z_PbPb_over_tot}
\end{figure}

Part of the uncertainties in Figure~\ref{Fig:Z_PbPb} are, however, rather related to the
overall normalization and not so much to the shape of the predicted spectra. In this
respect good observables to consider are the rapidity spectra
of Z and $W^\pm$ bosons in Pb+Pb collisions normalized to the integrated
cross-section,
\begin{equation}
\frac{d^2\sigma^{\rm PbPb}_{\rm Norm}(\sqrt{s_{\rm PbPb}})}{dM^2dy_R}
\equiv
 \frac{1}{\sigma_{\rm tot}^{\rm PbPb}(\sqrt{s_{\rm PbPb}})} \frac{d^2\sigma^{\rm PbPb}(\sqrt{s_{\rm PbPb}})}{dM^2dy_R},
 \label{eq:PbPb_Ratio_tot}
\end{equation}
where $\sigma_{\rm tot}^{\rm PbPb}$ refers to the cross-section integrated over the rapidity.
These are shown in Figure~\ref{Fig:Z_PbPb_over_tot}. These quantities are good in the sense that they
are free from any overall normalization uncertainties i.e. can be formed equally from the
absolute yields ${d^2N^{\rm PbPb}}/{dM^2dy_R}$ even if the nucleon-nucleon luminosity is 
not known accurately. As can be seen by comparing Figures~\ref{Fig:Z_PbPb} and \ref{Fig:Z_PbPb_over_tot},
especially the free proton uncertainties get suppressed in the latter one. As the nuclear
modifications to the PDFs affect more the shape and not much the overall
magnitude of the spectra, the nuclear uncertainties remain essentially unchanged.

Apart from serving as tests of factorization and luminosity determination in PbPb collisions,
the quantities plotted in Figures~\ref{Fig:Z_PbPb} and \ref{Fig:Z_PbPb_over_tot} are not
well-suited to offer further constraints for the nuclear modifications of the PDFs: Even if the 
experimental errors were smaller than the uncertainty in Figures~\ref{Fig:Z_PbPb} and \ref{Fig:Z_PbPb_over_tot},
the fact that the uncertainties stemming from the free proton PDFs and those originating
from their nuclear modifications are of the same order makes it difficult to 
disentangle between the two.

\begin{figure}[!htb]
\center
\includegraphics[scale=0.25]{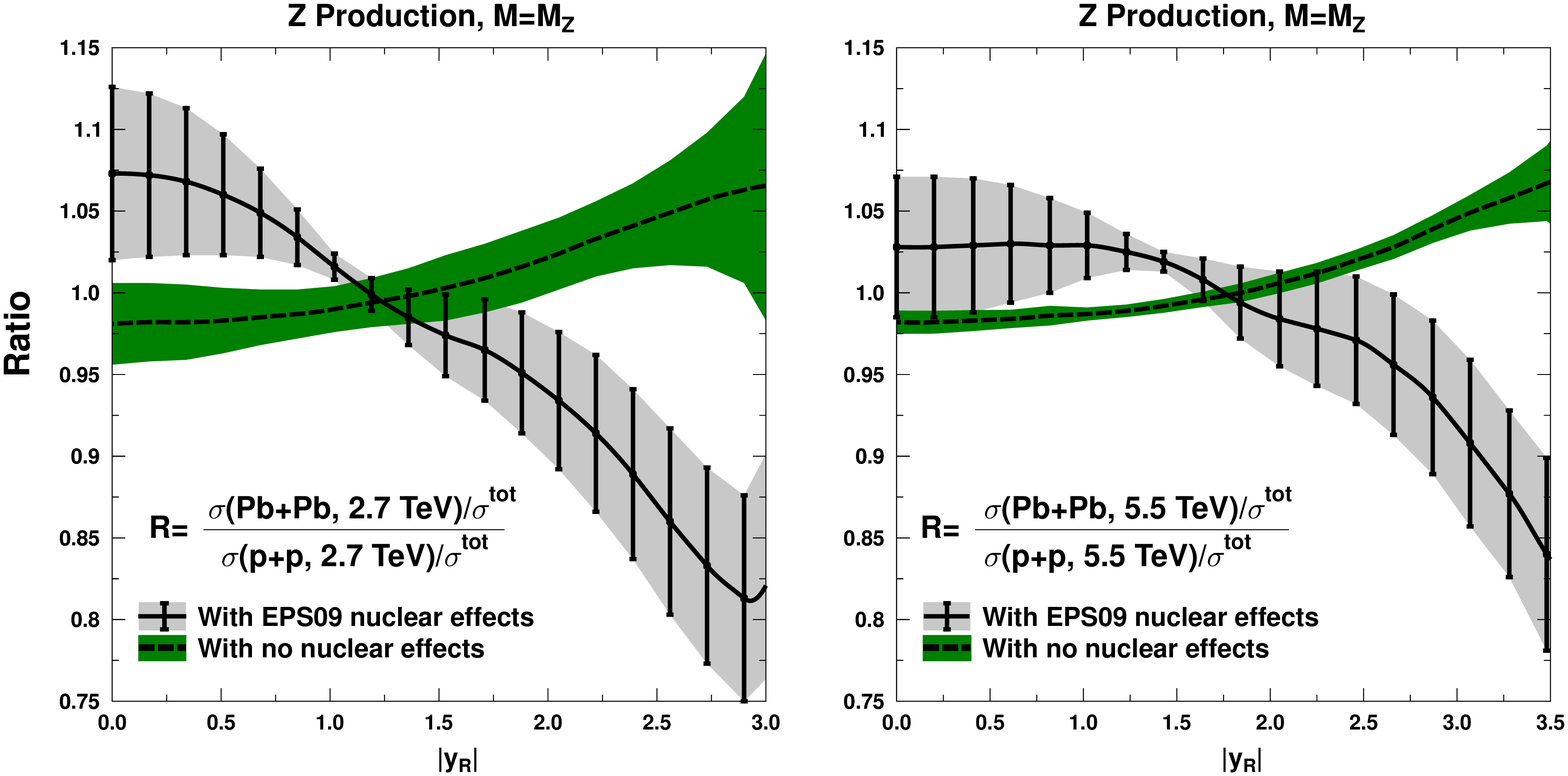}
\includegraphics[scale=0.25]{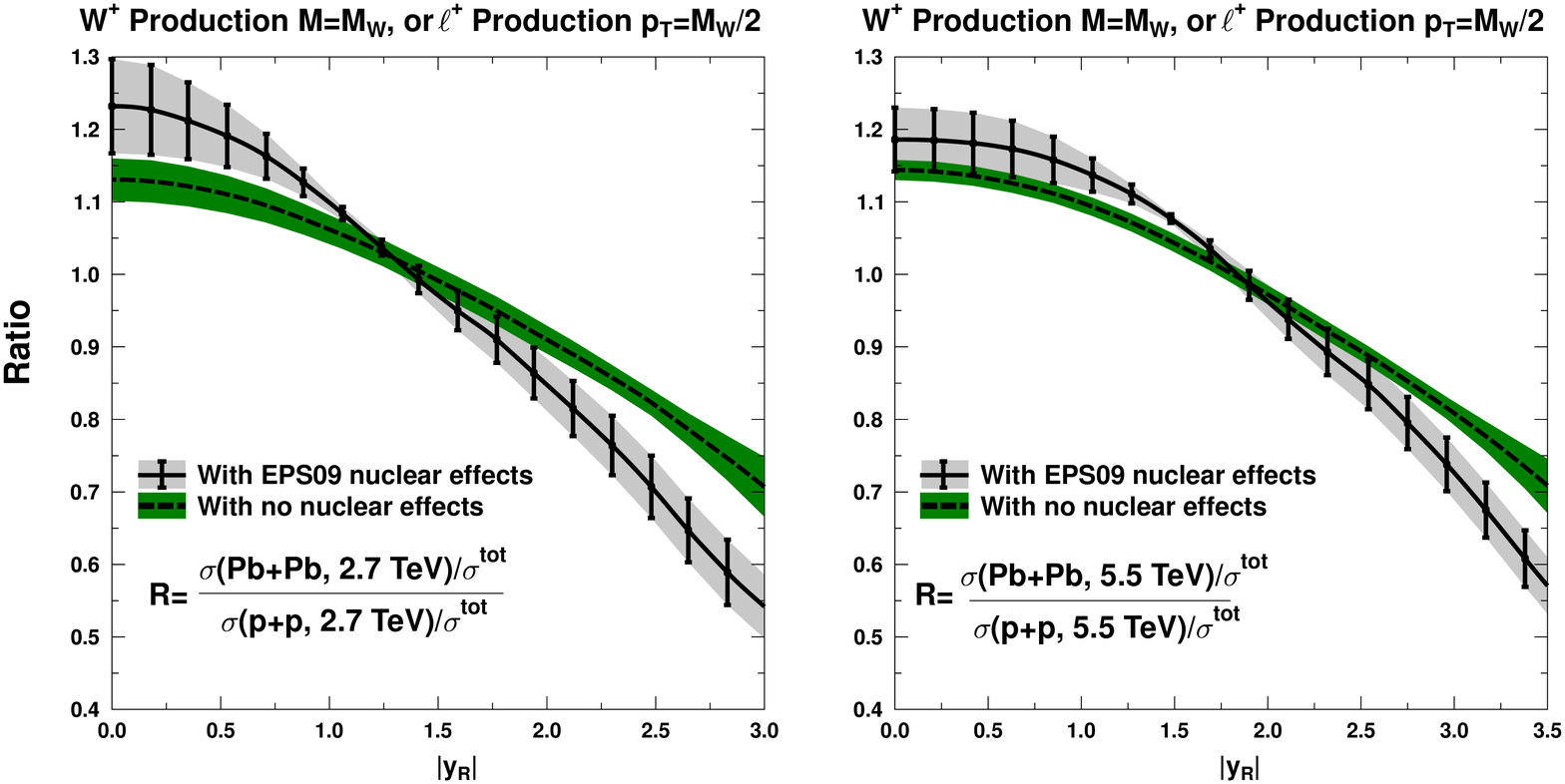}
\includegraphics[scale=0.25]{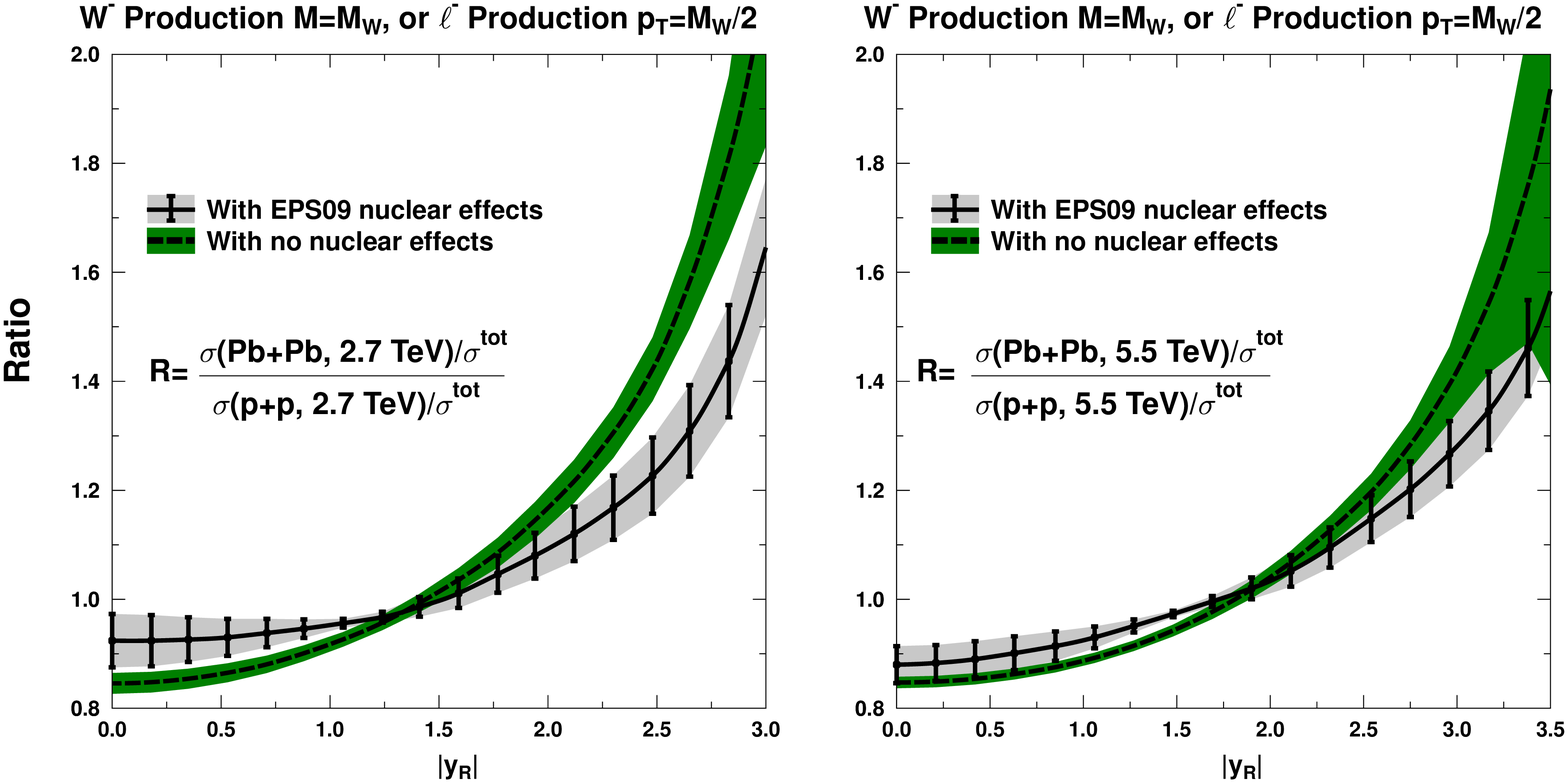}
\caption[]{\small
The predicted ratio between the normalized Z and W$^\pm$ cross-sections in PbPb and pp collisions at the heavy
boson pole, $M^2=M_{Z,W^\pm}^2$ ($p_T=M_W/2$ for charged lepton production) with $\sqrt{s}=2.7\,{\rm TeV}$ and $\sqrt{s}=5.5\,{\rm TeV}$. The green solid band represents the prediction calculated with CTEQ6.6 without applying the nuclear effects, and the solid black barred line with gray shade is the prediction computed by CTEQ6.6 applying the nuclear effects from EPS09. The error bars quantify the uncertainties resulting from the EPS09 uncertainty sets. The results are obtained by first calculating Eq.~(\ref{eq:PbPB_Ratio}) and normalizing it then by Eq.~(\ref{eq:PbPB_Ratio_Norm}). The symbol $\sigma^{\rm tot}$ refers to the cross-section integrated over the rapidity.}
\label{Fig:Z_PbPb_over_tot2}
\end{figure}

In order to further reduce the uncertainties due to the free proton PDFs and enhance
the visibility of the nuclear modifications, reference cross-sections
from pp-collisions would be needed. Avoiding the problems from the absolute the 
normalization uncertainty we look at ratios 
\begin{equation}
\frac{d^2\sigma^{\rm PbPb}_{\rm Norm}(\sqrt{s_{\rm PbPb}})}{dM^2dy_R} 
/ \frac{d^2\sigma^{\rm pp}_{\rm Norm}(\sqrt{s_{\rm pp}})}{dM^2dy_R}.
\label{eq:PbPB_Ratio}
\end{equation}
The problem here is, however, that the center-of-mass energies $\sqrt{s_{\rm pp}}$ (here we use $\sqrt{s_{\rm pp}}= 7 \, {\rm TeV}$) and $\sqrt{s_{\rm PbPb}}$ in pp- and PbPb-collisions will be non-equal at LHC, and the uncertainties from the free proton PDFs will not therefore perfectly cancel in these ratios. Rather than plotting directly
Eq.~(\ref{eq:PbPB_Ratio}), we normalize it by the factor
\begin{equation}
  \frac{d^2\sigma^{\rm pp}_{\rm Norm}(\sqrt{s_{\rm pp}})}{dM^2dy_R} 
/ \frac{d^2\sigma^{\rm pp}_{\rm Norm}(\sqrt{s_{\rm PbPb}})}{dM^2dy_R} , 
\label{eq:PbPB_Ratio_Norm}
\end{equation}
calculated with the central CTEQ6.6 set. This enhances the visibility of the 
nuclear effects eliminating the disbalance 
between the center-of-mass energies in pp and PbPb collisions. In other words, we calculate
\begin{equation}
R_{\rm PbPb} \equiv \frac{d^2\sigma^{\rm PbPb}_{\rm Norm}(\sqrt{s_{\rm PbPb}})}{dM^2dy_R} 
/  \frac{d^2\sigma^{\rm pp}_{\rm Norm}(\sqrt{s_{\rm PbPb}})}{dM^2dy_R} , 
 \label{eq:RPbPb}
\end{equation}
with the relative uncertainties from Eq.~(\ref{eq:PbPB_Ratio}). The results obtained in this way are plotted in Figure~\ref{Fig:Z_PbPb_over_tot2}. 

Interestingly, in the case of Z bosons the nuclear modifications cause the rapidity slope 
to change the sign with both center-of-mass energies. For W$^\pm$s the effects are
not that dramatic --- the deviations from unity are mainly caused by the isospin effects.
Although the free proton uncertainties are, on average, smaller than those from the EPS09,
they are in various places still too large to be neglected in a parton fit. Thus, we are
forced to conclude that --- in contrast what was observed about the p+Pb collisions --- the
 clean extraction of the PDF nuclear effects from Pb+Pb collisions
seems difficult. However, one should keep in mind that once the experimental data from the pp-runs
are included to the free proton PDF fits, one should observe a decrease in the free proton uncertainties
thereby increasing the usefulness of $R_{\rm PbPb}$s presented in Figure~\ref{Fig:Z_PbPb_over_tot2}.

\section{Conclusion}
\label{Conclusion}

We have reported a study of production of the heavy vector bosons (precisely to their leptonic decay channels)
in pPb and PbPb collisions at LHC.

From the pPb-study we find that these collisions at LHC would serve as a powerful
tool for testing the factorization and improving the present knowledge of the nuclear modifications
to the free nucleon PDFs. Especially, there would be no need  to use baseline pp data, but the nuclear
modifications could be extracted from the pPb data alone with very little sensitivity to the
underlying free proton PDFs utilizing the cross-section ratios
$$
\frac{d^2\sigma^{Z, y_R}_{\rm pPb}}{dM^2dy_R} /
\frac{d^2\sigma^{Z, -y_R}_{\rm pPb}}{dM^2dy_R} 
$$
and
$$
\left[ \frac{d^2\sigma^{W^+,y_R}}{dM^2dy_R} - \frac{d^2\sigma^{W^+, -y_R}}{dM^2dy_R} \right] / \left[ \frac{d^2\sigma^{W^-, y_R}}{dM^2dy_R} - \frac{d^2\sigma^{W^-, -y_R}}{dM^2dy_R} \right].
$$
The former quantity can currently be estimated with an approximate uncertainty of 10\%
outside the midrapidity ($y_R \gtrsim 1$), with uncertainty diminishing from that towards
$y_R = 0$. The latter quantity above carries around 20\% uncertainty in the whole rapidity
interval. The experimental data with errors smaller than these, would certainly offer
a possibility to test the factorization in nuclear collisions and to constrain the
nuclear modifications of the PDFs. It is worth stressing that these quantities do not suffer
from any uncertainties associated to the use of the Glauber model for normalization of the nuclear cross-sections. 

In contrast, distilling information about the nuclear effects in PDFs from symmetric PbPb collisions suffers
from a sensitivity to the the underlying free proton PDFs and their uncertainties. Moreover,
the predicted uncertainties from the nuclear corrections to the PDFs are not very
sizable. Therefore, the data from PbPb collisions will rather serve as a test of the
factorization and check of the Glauber model in PbPb collisions than as a constraint for the nPDFs.

\vspace{-0.3cm}
\section*{Acknowledgments}
CAS thanks Federico Antinori, David D'Enterria and John Jowett for useful discussions. 
This work is supported by Ministerio de Ciencia e Innovaci\'on of Spain under project FPA2009-06867-E; by Xunta de Galicia (Conseller\'{\i}a de Educaci\'on and Conseller\'\i a de Innovaci\'on e Industria -- Programa Incite); by the Spanish Consolider-Ingenio 2010 Programme CPAN (CSD2007-00042); and by by the European Commission grant
PERG02-GA-2007-224770. CAS is a Ram\'on y Cajal researcher.

\newpage
\begin{flushleft}
{\LARGE \bf Appendix}
\end{flushleft}

\appendix

\section*{A relation between $\ell^\pm$ and $W^\pm$ spectra}
\label{sec:Relation}

\begin{figure}[!htb]
\center
\vspace{-0.5cm}
\includegraphics[scale=0.35]{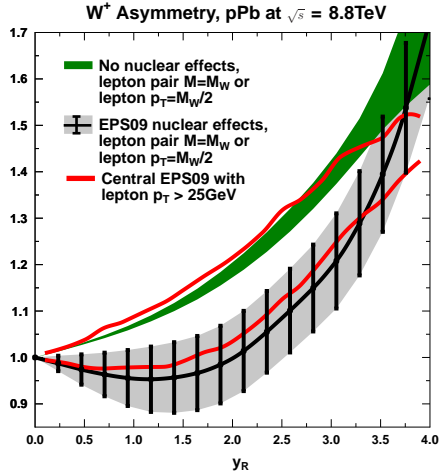}
\caption[]{\small
As the right panel of Figure~\ref{Fig:W_pPb_Asym1}, but the red lines indicating
$\sigma^{\ell^+, y_R} / \sigma^{\ell^+, -y_R}$
 computed at NLO with the central EPS09 set and $p_T>25 \, {\rm GeV}$.}
\label{Fig:W_pPb_Asym3}
\end{figure}

In the leading order, the cross-section for charged lepton production
can be written as
\begin{eqnarray}
E\frac{d^3\sigma^{H_1H_2 \rightarrow \ell^\pm + X}}{d^3p} & = & \frac{\alpha^2_{\rm em}}{\sin^4\theta_{\rm W}}
\sum_{i,j} |V_{ij}|^2
\int_{x_2^{\rm min}}^1 dx_2 \left( x_2 - \frac{p_T}{\sqrt{s}}e^{-y_R}\right)^{-1} \frac{x_1x_2}{6{\hat s}^2} \\
& & \hspace{-2cm} \times \frac{1}{(\hat{s}-M_W^2)^2 + M_W^2\Gamma_W^2} \left[{\hat t}^2q_i^{(1)}(x_1)\overline{q}_j^{(2)}(x_2) +
{\hat u}^2 \overline{q}_j^{(1)}(x_1){q}_i^{(2)}(x_2) \right] \nonumber
\end{eqnarray}
with
$$
\begin{array}{llc}
i = u, c     & j = d, s, b & {\rm for \,\, \ell^+} \\ 
i = d, s, b  & j = u, c    & {\rm for \,\, \ell^-},
\end{array}
$$
and
\begin{equation}
 \hat s = x_1x_2s, \quad \hat t = -\sqrt{s}p_Tx_1e^{-y_R}, \quad \hat u = -\sqrt{s}p_Tx_2e^{y_R}
\end{equation}
\begin{equation}
x_1 = \frac{x_2p_{T}e^{y_R}}{x_2\sqrt{s}-p_Te^{-y_R}}, \quad x_2^{\rm min} = \frac{p_{T}e^{-y_R}}{\sqrt{s}-p_Te^{y_R}}.
\end{equation}
In the narrow width approximation the cross-section is dominated
by the peak of the $W$-propagator, that is, when the kinematical configuration is
such that $\hat{s}-M_W^2 = 0$. At $p_T=M_W/2$ one easily finds that this happens when
\begin{equation}
 x_1 = \frac{M_W}{\sqrt{s}}e^{y_R}, \quad x_2 = \frac{M_W}{\sqrt{s}}e^{-y_R}, \quad \hat{t}=\hat{u} = -\frac{M_W^2}{2},
\end{equation}
so that
\begin{equation}
E\frac{d^3\sigma^{h_1h_2 \rightarrow \ell^\pm + X}}{d^3p}_{\Big| p_T = \frac{M_W}{2}} \propto  \sum_{i,j} |V_{ij}|^2 
\left[q_i^{(1)}(x_1)\overline{q}_j^{(2)}(x_2) + \overline{q}_j^{(1)}(x_1){q}_i^{(2)}(x_2) \right].
\end{equation}
This is the same partonic combination at the same values of $x_1$ and $x_2$ that enters also
to the leading order $W^\pm$ production discusssed in Section \ref{sec:W-Production}. Therefore,
the ratios of charged lepton production at $y_R^{\ell^\pm}$ and $p_T=M_W/2$, are very close
to those of $W^\pm$ production at $y_R^{W^\pm}$ and $M=M_W$. Even in a wider leptonic $p_T$-window
the main results remain qualitatively unchanged, although the effects would
naturally slightly depend on the actual value of the cuts. As an example,
Figure~\ref{Fig:W_pPb_Asym3} shows how much difference would be induced in
$\sigma^{\ell^+, y_R} / \sigma^{\ell^+, -y_R}$ if $p_T>25 \, {\rm GeV}$ is
used instead of $p_T=M_W/2$. For these $p_T$-integrated results, the MCFM
code \cite{Campbell:1999ah} has been used.

\end{document}